\documentclass[a4paper,twoside]{article}

\usepackage{epsfig}
\usepackage{graphicx}
\usepackage{calc}
\usepackage{amssymb}
\usepackage{amstext}
\usepackage{amsmath}
\usepackage{amsthm}
\usepackage{color}
\usepackage{url}
\usepackage{multicol}
\usepackage{pslatex}
\usepackage{multirow}
\usepackage{apalike}
\usepackage{array}

\usepackage{SCITEPRESS}     
\usepackage{mathtools}

\begin{document}

\def\N{{\mathbb N}}
\def\Z{{\mathbb Z}}
\def\R{{\mathbb R}}

\newcommand{\YF}[1]{\begin{color}{red}\textit{#1}\end{color}}

\DeclarePairedDelimiter\abs{\lvert}{\rvert}%
\DeclarePairedDelimiter\norm{\lVert}{\rVert}%

\makeatletter
\let\oldabs\abs
\def\abs{\@ifstar{\oldabs}{\oldabs*}}
\let\oldnorm\norm
\def\norm{\@ifstar{\oldnorm}{\oldnorm*}}
\makeatother

\def\rl{{^{.}}}
\renewcommand\vec[1]{\overrightarrow{#1}}

\title{A Second Order Derivatives based Approach for Steganography}

\author{
\authorname{Jean-Fran\c{c}ois Couchot\sup{1}, 
Rapha\"el Couturier\sup{1},
Yousra Ahmed Fadil\sup{1,2},
Christophe Guyeux\sup{1}}
\affiliation{\sup{1}FEMTO-ST Institute, University of Franche-Comté, Rue du Maréchal Juin, Belfort, France}
\affiliation{\sup{2}College of engineering, University of Diyala, Baqubah, Iraq}
\email{\{jean-francois.couchot, raphael.couturier, yousra\_ahmed.fadil, christophe.guyeux\}@univ-fcomte.fr}
}


\abstract{
Steganography schemes are designed with the objective of minimizing a defined distortion function. In most existing state of the art approaches,
this distortion function is based on image feature preservation.
Since smooth regions or clean edges define image core, even a small 
modification in these areas largely modifies image features and is thus easily 
detectable. 
On the contrary, textures, noisy or chaotic regions are so difficult to model that the features having been modified inside these areas are similar to the initial ones.
These regions are characterized by disturbed level curves.
This work presents a new distortion function for steganography that is 
based on second order derivatives, which  are mathematical tools that
usually evaluate level curves. Two methods are explained to compute these 
partial derivatives and have been completely implemented. The first experiments
show that these approaches are promising.}

\onecolumn \maketitle \normalsize \vfill

\section{\uppercase{Introduction}}
\label{sec:introduction}

The objective of any \emph{steganographic} approach is to dissimulate a message into another one in an imperceptible way. In the context of this work, the host message is an image in the spatial domain, \textit{e.g.}, a raw image. 
A coarse steganographic technique consists in replacing the Least Significant Bit (LSB) of each pixel with the bits of the message to hide.
On the contrary, the goal of a \emph{steganalysis} approach is to decide whether a given content embeds or not a hidden message.

Steganographic schemes are evaluated according to their ability to face steganalyser tools. 
The efficiency of the former increases with the number of errors produced by the latter.
An error  is either a false positive decision or a  false negative one.
In the former case, the image is abusively declared to contain a hidden message 
whereas it is an original host.
In the latter case, the image is abusively declared as free of hidden content 
while it embeds a message.
The average error is thus the mean of these two ones.   
Let us select a security level expressed as a number in $[0,0.5]$, 
when developing a new steganographic scheme, the objective is to find an approach that
maximizes the size of the message that can be embedded in any image
with an average error larger than 
this security level.

Creating an efficient steganographic scheme aims at designing
an accurate distortion function that associates to each pixel the ability of 
modifying it. This function indeed allows the extraction the set of pixels that can be modified
with the smallest detectability. 
Highly Undetectable steGO (HUGO)~\cite{DBLP:conf/ih/PevnyFB10},
WOW~\cite{conf/wifs/HolubF12}, UNIWARD~\cite{HFD14}, STABYLO~\cite{DBLP:journals/adt/CouchotCG15},
EAI-LSBM~\cite{5411758}, and MVG~\cite{FK2013} 
are some of the  most efficient instances of such schemes.
The next step, \textit{i.e.}, the embedding process, is often 
common to all the steganographic schemes. For instance, this final step is the Syndrome-Trellis Code (STC)~\cite{FillerJF11} in many steganographic schemes
like the aforementioned ones.

The distortion function of HUGO evaluates for each
pixel in $(x,y)$ the sum of the directional SPAM features of the cover and of the image after modifying its
value $P(x,y)$.
In STABYLO and EAI-LSBM, the distortion functions are based on edge detection. The higher the difference between two consecutive pixels is, the smaller its distortion value is.
WOW (and similarly UNIWARD) distortion function is based on wavelet-based directional filters. 
These filters are applied twice to evaluate the cost of $\pm 1$ modification of the cover.
In all these previously detailed schemes, the function is 
designed to focus on a specific area, namely textured or noisy regions
where it is difficult to provide an accurate model.
The distortion function of MVG, for its part, is based on minimizing the 
Kullback-Leibler divergence.

In all aforementioned schemes, the distortion function returns a
large value in a easy-modelable smooth area and
a small one  in  textured, a "chaotic" area, \textit{i.e.},
where there is no model.
In other words, these approaches assign a large value to pixels  
that are in a specific level curve: modifying this pixel leads to associating another level to this pixel.
Conversely, when a pixel is not in a well defined level curve, its modification is hard to detect.

The mathematical tools that usually evaluate the level curves
are first and second order derivatives. Level curves are indeed defined
to be orthogonal to vectors of first order derivatives, \textit{i.e.}, to \emph{gradients}.
Second order derivatives allow to detect whether these level curves are locally 
well defined or, on the contrary, change depending on neighborhood.
Provided we succeed in defining a function $P$ 
that associates to each pixel $(x,y)$ its value $P(x,y)$,
pixels such that all the second order derivatives 
having high values are  good candidates to embed the message bits.

However, such a function $P$ is only known on pixels, \textit{i.e.}, on a finite set of points.
Its first and second derivatives cannot thus be mathematically computed. At most, one can provide 
approximate functions on the set of pixels. Even with such a function, ordering pixels according 
to the values of the Hessian matrix (\textit{i.e.}, the matrix of second order derivatives) 
is not a natural task.

This work first explains how such first and second order approximations can be computed on  
numerical images (Section~\ref{sec:gradient}).
Two proposals to compute second order derivatives are proposed and proven
(Section~\ref{sec:second} and Section~\ref{sec:poly}). This is the main contribution of this work. An adaptation of an existing distortion function is studied in Section~\ref{sec:distortion}. A whole set of experiments is presented in Section~\ref{sec:experiments}.
Concluding remarks and future work are presented in the last section.

\section{Derivatives in an Image}\label{sec:gradient}

This section first recalls links between level curves, gradient, and
Hessian matrix (Section~\ref{sub:general}).
It next analyses them using kernels from signal theory 
(Section~\ref{sub:class:1} and Section~\ref{sub:class:2}).

\subsection{Hessian Matrix}\label{sub:general}
Let us consider that an image can be seen as a numerical function $P$ that 
associates a value $P(x,y)$ to each pixel of coordinates $(x,y)$.
The variations of this function in $(x_0,y_0)$ 
can be evaluated thanks to its gradient
$\nabla{P}$, which is the vector whose two components 
are the partial derivatives in $x$ and in $y$ of $P$: 

\[\nabla{P}(x_0,y_0) = \left(\frac{\partial P}{\partial x}(x_0,y_0),\frac{\partial P}{\partial y}(x_0,y_0)\right).
\]

In the context of two variables, the gradient vector
points to the direction where the function has the highest increase.
Pixels with close values thus follow level curve that is orthogonal 
to the one of highest increase.

The variations of the gradient vector are expressed in the 
Hessian matrix $H$ of second-order partial derivatives of $P$.

\[
H = \begin{bmatrix} 
\dfrac{\partial^2 P}{\partial x^2} & 
\dfrac{\partial^2 P}{\partial x \partial y} \\
\dfrac{\partial^2 P}{\partial y \partial x} & 
\dfrac{\partial^2 P}{\partial y^2} \\
\end{bmatrix}. 
\]

In one pixel $(x_0,y_0)$, the larger the absolute values of this matrix are,
the more the gradient is varying around $(x_0,y_0)$.
We are then left to evaluate such an Hessian matrix.

This task is not as easy as it appears since natural images are not defined 
with differentiable functions from $\R^2$ to $\R$.
Following subsections provide various approaches to compute these 
Hessian matrices.

\subsection{Classical Gradient Image Approaches}\label{sub:class:1}
In the context of image values, the most used approaches to evaluate gradient vectors are the well-known ``Sobel'', ``Prewitt'', ``Central Difference'', and ``Intermediate Difference'' ones.

\begin{table}[ht]
\caption{Kernels of usual image gradient operators\label{table:kernels:usual}
}
\centering
\scriptsize
\begin{tabular}{|c|c|c|}
    \hline
    Name&   Sobel & Prewitt \\
    \hline
    Kernel & $\textit{Ks}= \begin{bmatrix} -1 & 0 & +1 \\ -2 & 0 & +2 \\ -1 & 0 & +1 \end{bmatrix} $ &
    $\textit{Kp}= \begin{bmatrix} -1 & 0 & +1 \\ -1 & 0 & +1 \\ -1 & 0 & +1 \end{bmatrix} $\\
    \hline\hline
    Name & Central & Intermediate \\
            & Difference &Difference \\ 
    \hline
    Kernel & $\textit{Kc}= \begin{bmatrix} 0&0&0 \\ -\dfrac{1}{2} & 0 & +\dfrac{1}{2} \\ 0&0&0 \end{bmatrix} $ &
    $\textit{Ki}= \begin{bmatrix} 0&0&0 \\ 0 & -1 & 1 \\ 0&0&0 \end{bmatrix} $ \\
    \hline
\end{tabular}
\end{table}

Each of these approaches applies a convolution product $*$ between a  kernel $K$ (recalled in Table~\ref{table:kernels:usual}) and 
 a $3\times 3$ window of pixel values $A$. The result 
 $A * K$ is an evaluation of the horizontal gradient, 
\textit{i.e.}, $\dfrac{\partial P}{\partial x}$
expressed as a matrix in $\R$.
Let $K\rl$ be the result of a $\pi/2$ rotation applied on $K$. 
The vertical gradient $\dfrac{\partial P}{\partial y}$ is similarly obtained by computing $A * K\rl$, which is again expressed as a matrix in $\R$.

The two elements of the first line 
of the Hessian matrix are the result
of applying the horizontal gradient calculus 
first on  $\dfrac{\partial P}{\partial x}$ and next 
on  $\dfrac{\partial P}{\partial y}$. 
Let us study these Hessian matrices in the next section.

\subsection{Hessian Matrices induced by Gradient Image Approaches}\label{sub:class:2}

First of all, it is well known that 
$\dfrac{\partial^2 P}{\partial x \partial y} $ is equal to 
$\dfrac{\partial^2 P}{\partial y \partial x}$ if  
the approach that computes the gradient and the
one which evaluates the Hessian matrix are the same.
For instance, in the Sobel approach, 
it is easy to verify that the calculus of 
$\dfrac{\partial^2 P}{\partial x \partial y}$ 
and of 
$\dfrac{\partial^2 P}{\partial y \partial x}$ 
are both the result 
of a convolution product with the Kernel 
$\textit{Ks}''_{xy}$ given in Table~\ref{table:hessian:usual}.
This one summarizes kernels
$K_{x^2}''$ and 
$K_{xy}''$  
that allow to respectively compute  
$\dfrac{\partial^2 P}{\partial x^2}$ and
$\dfrac{\partial^2 P}{\partial x \partial y}$ with a convolution product
for each of the usual image gradient operator.

\begin{table}[ht]
\caption{Kernels of second order gradient operators\label{table:hessian:usual}
}\centering
\tiny
\begin{tabular}{|c|c|}
    \hline
    Sobel & Prewitt \\
    \hline
    $
    \textit{Ks}_{x^2}''= 
    \begin{bmatrix} 
    1 &  0  & -2  & 0 & 1 \\ 
    4 &  0  & -8  & 0 & 4 \\ 
    6 &  0  & -12 & 0 & 6 \\
    4 &  0  & -8  & 0 & 4 \\ 
    1 &  0  & -2  & 0 & 1 
    \end{bmatrix} 
    $
    & 
    $
    \textit{Kp}_{x^2}''= 
    \begin{bmatrix} 
    1 &  0  & -2  & 0 & 1 \\ 
    2 &  0  & -4  & 0 & 2 \\ 
    3 &  0  & -6 & 0 & 3 \\
    2 &  0  & -4  & 0 & 2 \\ 
    1 &  0  & -2  & 0 & 1 
    \end{bmatrix} 
    $
    \\
    \hline
    $
    \textit{Ks}_{xy}''=
    \begin{bmatrix} 
    -1 & -2 & 0 & 2 & 1 \\ 
    -2 & -4 & 0 & 4 & 2 \\ 
    0  & 0  & 0 & 0 & 0 \\
    2 & 4 & 0 & -4 & -2 \\ 
    1 & 2 & 0 & -2 & -1 
    \end{bmatrix} 
    $
    &
    $
    \textit{Kp}_{xy}''=
    \begin{bmatrix} 
    -1 & -1 & 0 & 1 & 1 \\ 
    -1 & -1 & 0 & 1 & 1 \\ 
    0  & 0  & 0 & 0 & 0 \\
    1 & 1 & 0 & -1 & -1 \\ 
    1 & 1 & 0 & -1 & -1 
    \end{bmatrix} 
    $ \\
    
    \hline\hline
    Central & Intermediate \\
    Difference &Difference \\ 
    \hline
    $
    \textit{Kc}_{x^2}''= 
    \begin{bmatrix} 
    0 &  0  & 0  & 0 & 0 \\ 
    0 &  0  & 0  & 0 & 0 \\ 
    \dfrac{1}{4} &  0  & -\dfrac{1}{2} & 0 & \dfrac{1}{4} \\
    0 &  0  & 0  & 0 & 0 \\ 
    0 &  0  & 0  & 0 & 0  
    \end{bmatrix} 
    $
    & 
    $
    \textit{Ki}_{x^2}''= 
    \begin{bmatrix} 
    0 &  0  & 0  & 0 & 0 \\ 
    0 &  0  & 0  & 0 & 0 \\ 
    0 &  0  & 1  & -2 & 1 \\
    0 &  0  & 0  & 0 & 0 \\ 
    0 &  0  & 0  & 0 & 0  
    \end{bmatrix} 
    $
    \\
    \hline
    $
    \textit{Kc}_{xy}''= 
    \begin{bmatrix} 
     -\dfrac{1}{4}  & 0 & \dfrac{1}{4}\\ 
     0  & 0 & 0 \\ 
     \dfrac{1}{4} & 0 &  -\dfrac{1}{4} 
    \end{bmatrix} 
    $
    & 
    $
    \textit{Ki}_{xy}''= 
    \begin{bmatrix} 
     0  & -1 & 1 \\ 
     0  & 1 & -1 \\ 
     0  &  0 & 0 
    \end{bmatrix} 
    $\\
    \hline
\end{tabular}

\end{table}

The Sobel kernel $\textit{Ks}_{x^2}''$ allows to detect whether the central 
pixel belongs to a ``vertical'' edge, even if this one is noisy, by considering its 
vertical neighbours. The introduction of these vertical neighbours in this kernel is meaningful 
in the context of finding edges, but not very accurate when the objective is to 
precisely find the level curves of the image. 
Moreover, all the pixels that are in the  second and the fourth column in this kernel
are ignored.
The Prewitt Kernel has similar drawbacks in this context.

The Central Difference kernel $\textit{Kc}_{x^2}''$ is not influenced by 
the vertical neighbours of the central pixel and is thus more accurate here.
However, the kernel $\textit{Kc}_{xy}''$ again looses the values of the pixels that 
are vertically and diagonally aligned with the central one.

Finally, the Intermediate Difference kernel $\textit{Ki}_{x^2}''$ shifts 
to the left the value of horizontal variations of $\dfrac{\partial P}{\partial x}$:
the central pixel $(0,0)$ exactly receives the value
$\dfrac{P(0,2)-P(0,1)}{1} - \dfrac{P(0,1)-P(0,0)}{1}$,
which is an approximation of 
$\dfrac{\partial P}{\partial x}(0,1)$ and not of 
$\dfrac{\partial P}{\partial x}(0,0)$.
Furthermore the Intermediate Difference kernel $\textit{Ki}_{xy}''$ only deals 
with pixels in the upper right corner, loosing all the other information.

Due to these drawbacks, we are then left to produce another approach to find the level curves with strong accuracy.

\section{Second Order Kernels for Accurate Level Curves}\label{sec:second}

This step aims at finding 
accurate level curve variations in an image. 
We do not restrict the kernel to have a fixed size (\textit{e.g.}, $3\times3$ or $5 \times 5$ as in the 
aforementioned schemes).
This step is thus defined with kernels of size 
$(2n+1)\times (2n+1)$, $n \in \{1,2,\dots,N\}$, where 
$N$ is a parameter of the approach.

The horizontal gradient variations are thus captured thanks to $(2n+1)\times (2n+1)$ square kernels
\begin{small}
\[
\arraycolsep=1.4pt
\def\arraystretch{1.4}
\def\arraystretch{1.4}
    \textit{Ky}_{x^2}''=
    \left(
    \begin{array}{ccccccccc}
         0           & & & & \dots& & & & 0  \\
         \vdots      & & & & & & & & \vdots \\
         0           & & & & \dots & & & & 0  \\
         \dfrac{1}{2n}& 0 & \dots & 0  & -\dfrac{2}{2n} & 0  & \dots & 0& \dfrac{1}{2n} \\
         0           & & & & \dots& & & & 0  \\
        \vdots      & & & & & & & & \vdots \\
         0           & & & & \dots & & & & 0  
    \end{array}
    \right)
\]
\end{small}

When the convolution product is applied on a $(2n+1)\times(2n+1)$ window,
the result is 
$\dfrac{1}{2}\left(\dfrac{P(0,n)-P(0,0)}{n} - \dfrac{P(0,0)-P(0,-n)}{n}\right)$, which is indeed 
the variation between the gradient around the central pixel. 
This proves that this calculus is a correct approximation of 
$\dfrac{\partial^2 P}{\partial x^2}$.

When $n$ is 1, this kernel is a centered version of the horizontal Intermediate Difference kernel $\textit{Ki}_{x^2}''$ modulo a multiplication by $1/2$. When $n$ is 2, this kernel is equal to $\textit{Kc}_{x^2}''$.

The vertical gradient variations are again obtained by applying 
a $\pi/2$ rotation to each horizontal kernel $\textit{Ky}_{x^2}''$.

The diagonal gradient variations are obtained thanks to the $(2n+1)\times (2n+1)$ square kernels 
$\textit{Ky}_{xy}''$ defined by

\begin{small}
\[
\arraycolsep=1.4pt
\def\arraystretch{1.4}
\textit{Ky}_{xy}'' = \dfrac{1}{4}
\left(
    \begin{array}{ccccccccc}
     \frac{1}{n^2}& \dots & \frac{1}{2n} & \frac{1}{n} 
     & 0 &
     -\frac{1}{n}&-\frac{1}{2n} & \dots &  -\frac{1}{n^2} 
     \\
     \vdots & 0     &    &
     & \dots &
      &  &  0 & \vdots 
     \\
     \frac{1}{2n} & 0 &        &
     & \dots &
        &  & 0& -\frac{1}{2n} 
     \\
     \frac{1}{n} & 0 &    &    
     & \dots &
      &  & 0 &  -\frac{1}{n}
     \\
     0      &      & & & \dots& & & & 0  \\
     -\frac{1}{n} & 0 &        &
     & \dots &
        &  &0 & \frac{1}{n}
     \\
     -\frac{1}{2n} & 0 &       &
     & \dots &
      &  & 0 &  \frac{1}{2n} 
     \\
      \vdots & 0 &    &    
     & \dots &
        &  & 0& \vdots 
     \\
     -\frac{1}{n^2}&  \dots & -\frac{1}{2n} & -\frac{1}{n} 
     & 0 &
     \frac{1}{n}& \frac{1}{2n} & \dots  & \frac{1}{n^2}
    \end{array}
    \right).
\]
\end{small}

When $n$ is 1, $\textit{Ky}_{xy}''$ is equal to the kernel
$\textit{Kc}_{xy}''$, and 
%
the average vertical variations of the horizontal variations are 
\[
\begin{array}{l}
\dfrac{1}{4}
\left[ 
\left((P(0,1)-P(0,0))-(P(1,1)-P(1,0))\right)+ \right.\\
\quad \left((P(-1,1)-P(-1,0))-(P(0,1)-P(0,0))\right)+\\
\quad \left((P(0,0)-P(0,-1))-(P(1,0)-P(1,-1))\right)+\\
\quad \left. \left((P(-1,0)-P(-1,-1))-(P(0,0)-P(0,-1))\right)
\right]  \\
=\dfrac{1}{4}
\left[ P(1,-1) -P(1,1) - P(-1,-1) + P(-1,1)\right].
\end{array}
\]
which is $\textit{Ky}_{xy}''$.

Let us now consider any number $n$, $1 \le n \le N$.
Let us first investigate the vertical variations related to 
the horizontal vector $\vec{P_{0,0}P_{0,1}}$ 
(respectively  $\vec{P_{0,-1}P_{0,0}}$)
of length 1 that starts from (resp. that points to) $(0,0)$. 
As with the case $n=1$, there are 2 new vectors of 
length 1, namely 
$\vec{P_{n,0}P_{n,1}}$ and $\vec{P_{-n,0}P_{-n,1}}$ 
(resp. 
$\vec{P_{n,-1}P_{n,0}}$, and $\vec{P_{-n,-1}P_{-n,0}}$)
that are vertically aligned with $\vec{P_{0,0}P_{0,1}}$ 
(resp. with $\vec{P_{0,-1}P_{0,0}}$).

The vertical variation is now equal to $n$. Following the case where $n$ is 1 to compute the average variation, 
the coefficients of the first and last line around the central 
vertical line are thus from left to right:
$\dfrac{1}{4n}$,
$\dfrac{-1}{4n}$,
$\dfrac{-1}{4n}$, and
$\dfrac{1}{4n}$.

Cases are similar with vectors $\vec{P_{0,0}P_{0,1}}$, \ldots  
$\vec{P_{0,0}P_{0,n}}$ which respectively lead to coefficients 
$-\dfrac{1}{4 \times 2n}$, \ldots, 
$-\dfrac{1}{4 \times n.n}$, and the proof is omitted.
Finally, let us consider the vector $\vec{P_{0,0}P_{0,1}}$
and its vertical variations when $\delta y$ is $n-1$.
As in the case where $n=1$, we thus obtain the coefficients 
$\dfrac{1}{4 \times (n-1)n}$ and 
$-\dfrac{1}{4 \times (n-1)n}$ 
 (resp. $-\dfrac{1}{4 \times (n-1)n}$ and 
$\dfrac{1}{4 \times (n-1)n}$)
in the second line (resp. in the 
penultimate line) since the vector has length $n$
and $\delta y$ is $n-1$.
Coefficient in the other lines are similarly obtained and the proof is thus omitted.

We are then left to compute an approximation of the partial second order derivatives 
$\dfrac{\partial^2 P}{\partial x^2}$,  $\dfrac{\partial^2 P}{\partial y^2}$, and $\dfrac{\partial^2 P}{\partial x \partial y}$
with the kernels, 
$\textit{Ky}_{x^2}''$, $\textit{Ky}_{y^2}''$, and $\textit{Ky}_{xy}''$ respectively.
However, the size of each of these kernels is varying from $3\times3$ to $(2N+1)\times (2N+1)$.
Let us explain the approach on the former partial derivative.
The other can be immediately deduced.

Since the objective is to detect large variations, the second order derivative is approximated as 
the maximum of the approximations. More formally, 
let $n$, $1 \le n \le N$, be an integer number and 
$\dfrac{\partial^2 P}{\partial x^2}_n$ be the result of applying the Kernel 
$\textit{Ky}_{x^2}''$ of size $(2n+1)\times (2n+1)$. The derivative 
$\dfrac{\partial^2 P}{\partial x^2}$ is defined by

\begin{equation}
\dfrac{\partial^2 P}{\partial x^2}
= \max \left\{ 
\abs{\dfrac{\partial^2 P}{\partial x^2}_1}, \dots, \abs{\dfrac{\partial^2 P}{\partial x^2}_N}
\right \}. 
\label{eq:d2p_dx2}
\end{equation}

The same iterative approach is applied to compute approximations of
$\dfrac{\partial^2 P}{\partial y \partial x}$ 
and of 
$\dfrac{\partial^2 P}{\partial y^2}$.
Next section studies the suitability of approximating second order derivatives
when considering an image as a polynomial.

\section{Polynomial Interpolation of Images for Hessian Matrix Computation}\label{sec:poly}
Let $P(x,y)$ be the discrete value of the pixel $(x,y)$ in the image.
Let $n$, $1 \le n \le N$, be an integer such that the objective is to find a polynomial interpolation 
on the $(2n+1)\times(2n+1)$ window where the central pixel has index $(0,0)$.  
There exists an unique polynomial $L : \R\times \R \to \R$ of degree $(2n+1)\times(2n+1)$ defined  
such that $L(x,y)=P(x,y)$ for each pixel $(x,y)$ in this window.
Such a polynomial is defined by 
\begin{equation}
\begin{array}{l}
L(x,y) =  
\sum_{i=-n}^{n}
\sum_{j=-n}^{n} \\
\quad P(i,j)
\left( 
\prod_{\stackrel{-n\leq j'\leq n}{j'\neq j}}
\frac{x-j'}{i-j'}
\right)
\left( 
\prod_{\stackrel{-n\leq i'\leq n}{i'\neq i}}
\frac{x-i'}{i-i'}
\right)
\end{array}
\end{equation}

It is not hard to prove that the first order horizontal derivative of the polynomial $L(x,y)$ 
is
\begin{equation}
\begin{array}{l}
\dfrac{\partial L}{\partial x} =  
\sum_{i=-n}^{n}
\sum_{j=-n}^{n} 
P(i,j)
\left( 
\prod_{\stackrel{-n\leq j'\leq n}{j'\neq j}}
\frac{y-j'}{j-j'}
\right)\\
\quad
\left( 
\sum_{\stackrel{-n\leq i'\leq n}{i'\neq i}}
\frac{1}{i-i'}
\prod_{\stackrel{-n\leq i''\leq n}{i''\neq i,i'}}
\frac{x-i''}{i-i''}
\right)
\end{array}
\end{equation}
\noindent and thus to deduce that the
second order ones are

\begin{equation}
\begin{array}{l}
\dfrac{\partial^2 L}{\partial x^2} =  
\sum_{i=-n}^{n}
\sum_{j=-n}^{n} 
P(i,j)
\left( 
\prod_{\stackrel{-n\leq j'\leq n}{j'\neq j}}
\frac{y-j'}{j-j'}
\right)\\
\quad
\left( 
\sum_{\stackrel{-n\leq i'\leq n}{i'\neq i}}
\frac{1}{i-i'}
\sum_{\stackrel{-n\leq i''\leq n}{i''\neq i,i'}}
\frac{1}{i-i''}
\prod_{\stackrel{-n\leq i'''\leq n}{i'''\neq i,i',i''}}
\frac{x-i'''}{i-i'''}
\right)
\end{array}
\label{eq:deriv:poly:x2}
\end{equation}

\begin{equation}
\begin{array}{l}
\dfrac{\partial^2 L}{\partial y \partial x} =  
\sum_{i=-n}^{n}
P(i,j) \\
\quad
\left( 
\sum_{\stackrel{-n\leq j'\leq n}{j'\neq j}}
\frac{1}{j-j'}
\prod_{\stackrel{-n\leq j''\leq n}{j''\neq j, j'}}
\frac{y-j''}{j-j''}
\right)\\
\quad
\left( 
\sum_{\stackrel{-n\leq i'\leq n}{i'\neq i}}
\frac{1}{i-i'}
\prod_{\stackrel{-n\leq i''\leq n}{i''\neq i, i'}}
\frac{x-i''}{i-i''}
\right)
\end{array}
\label{eq:deriv:poly:yx}
\end{equation}

These second order derivatives  are computed for each moving 
window and are associated to the central pixel, \textit{i.e.}, to the pixel $(0,0)$ inside this one.

Let us first simplify $\dfrac{\partial^2 L}{\partial x^2}$ when $(x,y)=(0,0)$
defined in Equation~(\ref{eq:deriv:poly:x2}). If $j$ is not null, the index $j'$
is going to be null and the product  
$\left( 
\prod_{\stackrel{-n\leq j'\leq n}{j'\neq j}}
\frac{-j'}{j-j'}
\right)$ is null too. 
In this equation, we thus only consider $j=0$.
It is obvious that the product indexed with $j'$ is thus equal to 1.
This equation can thus be simplified in:

\begin{equation}
\begin{array}{l}
\dfrac{\partial^2 L}{\partial x^2} =  
\sum_{i=-n}^{n}
P(i,0)\\
\quad
\left( 
\sum_{\stackrel{-n\leq i'\leq n}{i'\neq i}}
\frac{1}{i-i'}
\sum_{\stackrel{-n\leq i''\leq n}{i''\neq i,i'}}
\frac{1}{i-i''}
\prod_{\stackrel{-n\leq i'''\leq n}{i'''\neq i,i',i''}}
\frac{i'''}{i'''-i}
\right)
\end{array}
\label{eq:deriv:poly:x2:simpl}
\end{equation}

and then in:

\begin{equation}
\begin{array}{l}
\dfrac{\partial^2 L}{\partial x^2} =  
\sum_{i=-n}^{n}
P(i,0)\\
\quad
\left( 
\sum_{\stackrel{-n\leq i' < i'' \le n}{i',i''\neq i}}
\frac{2}{(i-i')(i-i'')}
\prod_{\stackrel{-n\leq i'''\leq n}{i'''\neq i,i',i''}}
\frac{i'''}{i'''-i}
\right).
\end{array}
\label{eq:deriv:poly:x2:simpl:2}
\end{equation}

From this equation, the kernel allowing to evaluate horizontal 
second order derivatives 
can be computed for any $n$.
It is further denoted as $Ko''_{x^2}$. 
Instances of such matrix when $n=2$, $3$, and $4$
are given in Table~\ref{table:sod:hori:poly}.

\begin{table}[ht]
    \caption{Kernels $Ko''_{x^2}$ 
    for second order horizontal derivatives induced 
    by polynomial interpolation}
    \centering
    \scriptsize
\def\arraystretch{1.4}
    \begin{tabular}{|c|c|}
         \hline
         $n$ & $Ko''_{x^2}$ \\
         \hline
         $2$ & $\left[\dfrac{-1}{12}, \dfrac{4}{3} , \dfrac{-5}{2}, \dfrac{4}{3} \dfrac{-1}{12}\right]$ \\
         \hline
         $3$ & $\left[\dfrac{1}{90}, \dfrac{-3}{20}, \dfrac{3}{2}, \dfrac{-49}{18}, \dfrac{3}{2}, \dfrac{-3}{20}, \dfrac{1}{90}\right]$ \\
         \hline
         $4$ & $\left[\dfrac{-1}{560}, \dfrac{8}{315}, \dfrac{-1}{5}, \dfrac{8}{5}, \dfrac{-205}{72}, \dfrac{8}{5}, \dfrac{-1}{5}, \dfrac{8}{315}, \dfrac{-1}{560}\right]$\\
         \hline
    \end{tabular}

    \label{table:sod:hori:poly}
\end{table}

\begin{table}[ht]
\caption{Kernels for second order diagonal derivatives induced 
    by polynomial interpolation \label{table:sod:diag:poly}
}
\centering
\scriptsize
\def\arraystretch{1.5}
\begin{tabular}{|c|c|}
    \hline
    $n$ & $Ko''_{xy}$\\
    \hline
    2 & $
\begin{bmatrix} 
\dfrac{1}{4} & 0 &  \dfrac{-1}{4}\\
 0 & 0 &0\\
\dfrac{-1}{4} & 0 &  \dfrac{1}{4}\\
\end{bmatrix}
$
 \\   
 \hline
   3 &  
   $\begin{bmatrix} 
\dfrac{1}{144} & \dfrac{-1}{18} & 0 & \dfrac{1}{18} & \dfrac{-1}{144}\\
\dfrac{-1}{18} & \dfrac{4}{9} & 0 & \dfrac{-4}{9} & \dfrac{1}{18}\\
0 & 0 & 0 & 0 &0\\
\dfrac{1}{18} & \dfrac{-4}{9} & 0 & \dfrac{4}{9} & \dfrac{-1}{18}\\
\dfrac{-1}{144} & \dfrac{1}{18} & 0 & \dfrac{-1}{18} & \dfrac{1}{144}
\end{bmatrix}
$\\
  \hline
\end{tabular}
\end{table}

From Equation~(\ref{eq:deriv:poly:yx}), kernels allowing to evaluate diagonal 
second order derivatives (\textit{i.e.}, 
$\dfrac{\partial^2 L}{\partial y \partial x}$)  are computed.
They are denoted as $Ko''_{xy}$.
Table~\ref{table:sod:diag:poly} gives two examples of them when $n=1$ and $n=2$.
Notice that for $n=1$, the kernel $Ko''_{xy}$ is equal to $Kc''_{xy}$.

\section{Distortion Cost}\label{sec:distortion}

The distortion function has to associate to each pixel $(i,j)$ 
the cost $\rho_{ij}$ of its modification
by $\pm 1$.

The objective is to map a small value to a pixel when all its second order derivatives 
are high and a large value otherwise. 
In WOW and UNIWARD  the distortion function is based on the H\"older norm with
\[
\rho_{ij}^w = 
\left(
\abs{\xi_{ij}^h}^{p} +
\abs{\xi_{ij}^v}^{p} +
\abs{\xi_{ij}^d}^{p} 
\right)^{-\frac{1}{p}}
\]
where $p$ is a negative number and 
$\xi_{ij}^h$ (resp. $\xi_{ij}^v$ and $\xi_{ij}^d$)
represents the horizontal (resp. vertical and diagonal) suitability.
A small suitability in one direction means an inaccurate position to embed a message.

We propose here to adapt such a distortion cost as follows:
\[
\rho_{ij} = 
\left(
\abs{\dfrac{\partial^2 P}{\partial x^2}(i,j)} +
\abs{\dfrac{\partial^2 P}{\partial y^2}(i,j)} +
\abs{\dfrac{\partial^2 P}{\partial y \partial x}(i,j)}
\right)^{-\frac{1}{p}}
\]
It is not hard to check that such a function has large 
value when at least one of its derivatives is null. Otherwise,
the larger the derivatives are, the smaller the returned value
is.

\begin{figure*}
    \centering
    \begin{tabular}{|c|c|c|}
    \hline
    Scheme & Stego. content & Changes with cover \\
    \hline
    $Ky$ based approach &\includegraphics[scale=0.20]{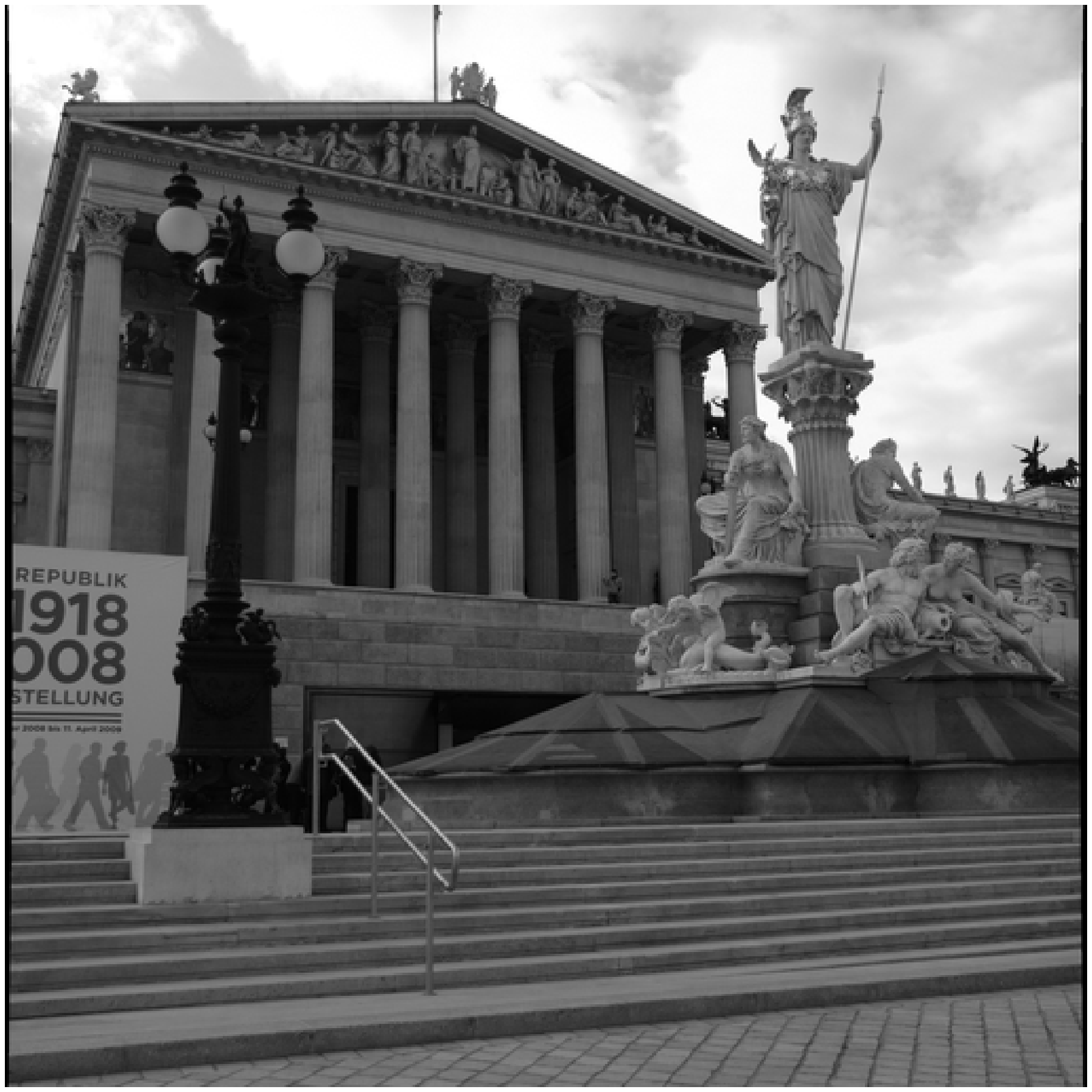}&
    \includegraphics[scale=0.20]{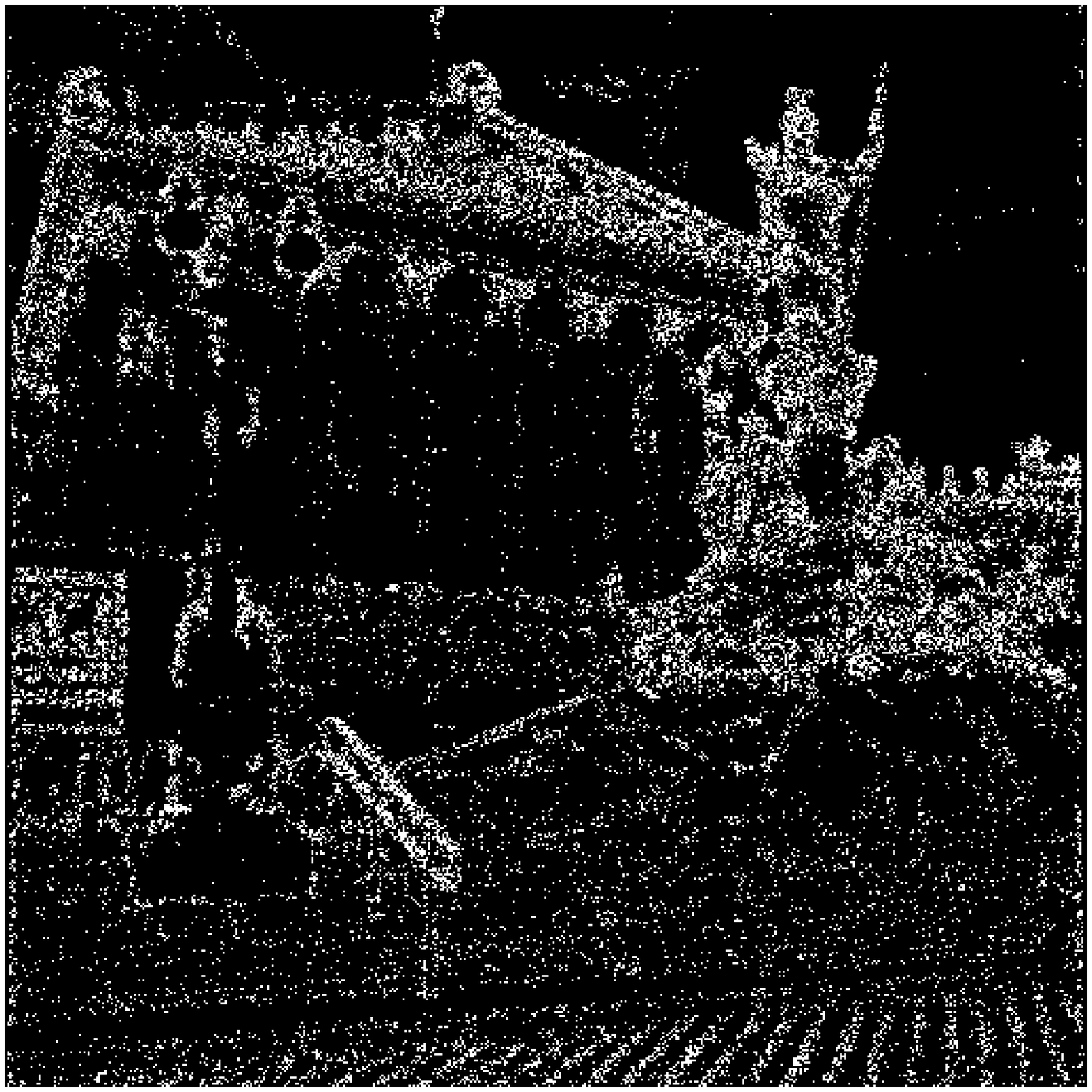} \\
    \hline
    $Ko$ based approach & \includegraphics[scale=0.20]{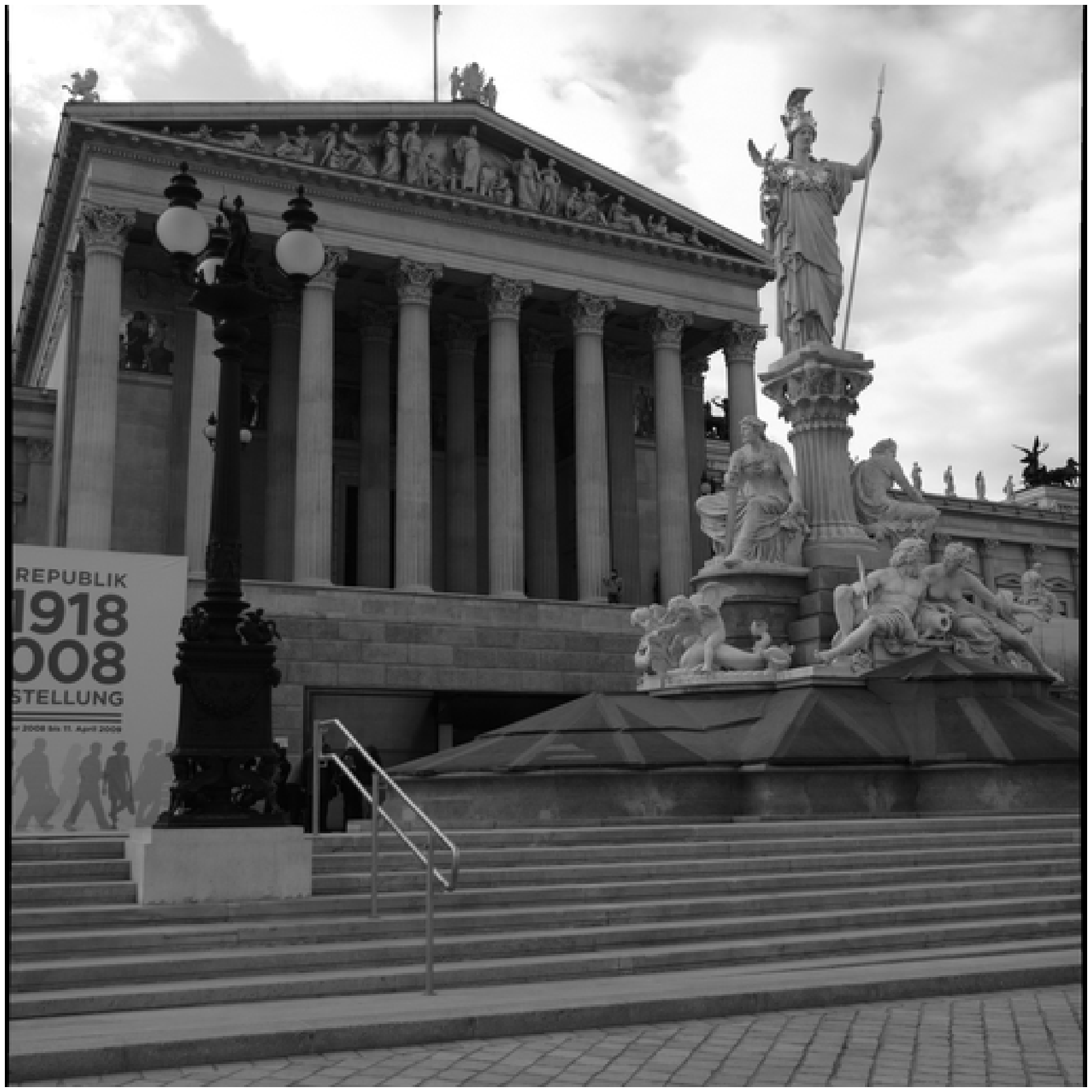} &
    \includegraphics[scale=0.20]{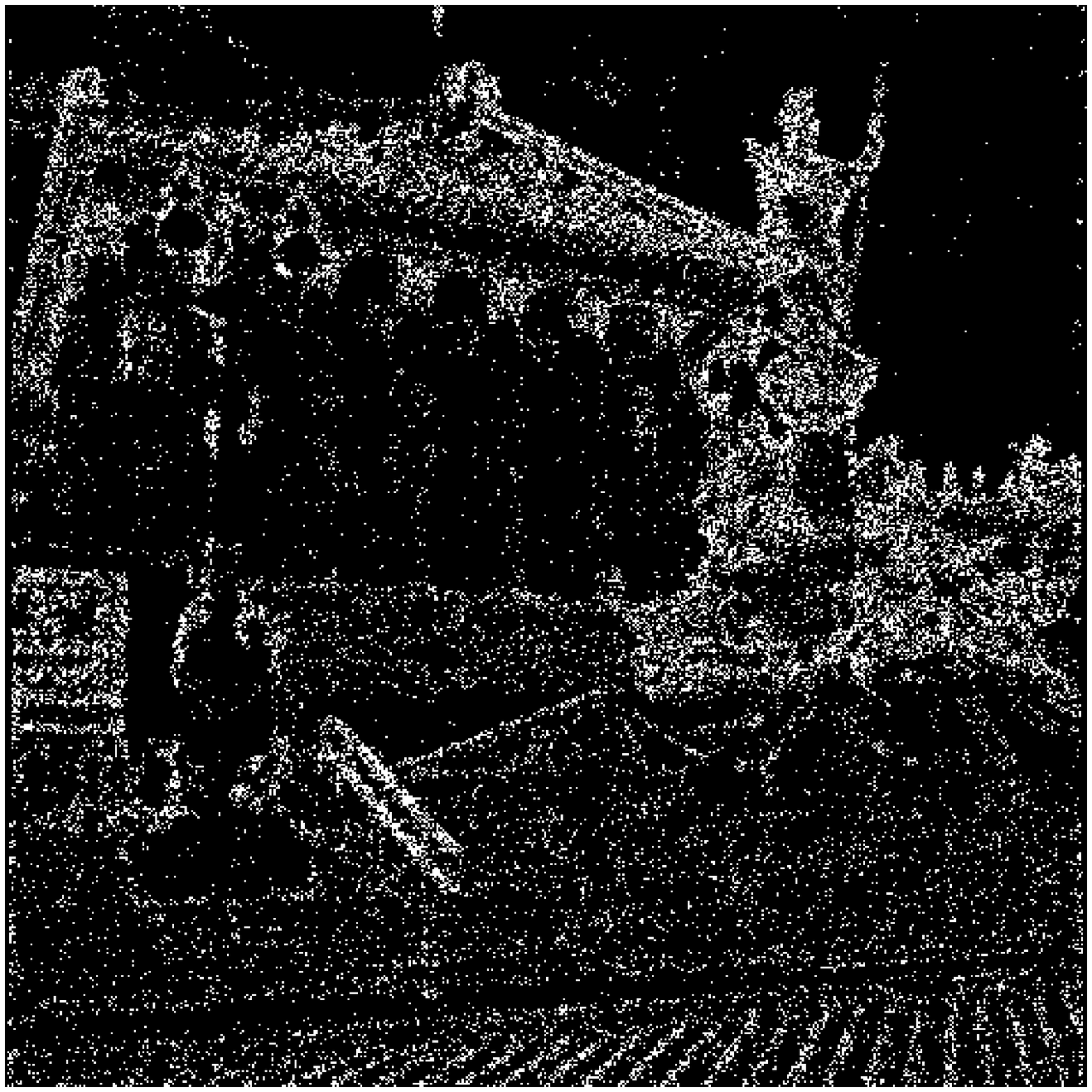} \\
    \hline
    \end{tabular}
    \caption{Embedding changes instance with payload  $\alpha = 0.4$}
    \label{fig:oneimage}
\end{figure*}

\section{Experiments}\label{sec:experiments}

First of all, the whole steganographic approach code is available online\footnote{\url{https://github.com/stego-content/SOS}}.

Figure~\ref{fig:oneimage} presents the results of embedding data in a cover image
from the BOSS contest database~\cite{Boss10} with respect to the two second order derivative schemes presented in this work. 
The $Ky$ based approach (resp. the $Ko$ based one)
corresponds to the scheme detailed in Section~\ref{sec:second} 
(resp. in Section~\ref{sec:poly}).
The payload $\alpha$ is set to 0.4 and kernels are computed with $N=4$.    
The central column outputs the embedding result whereas the right one displays 
differences between the cover image and the stego one.
It can be observed that pixels in smooth area (the sky, the external access steps) and pixels in clean edges (the columns, 
the step borders) are not modified by the approach.
On the contrary, an unpredictable area (a monument for example) concentrates pixel changes.

\subsection{Choice of parameters}

The two methods proposed in Section~\ref{sec:second} and  
in Section~\ref{sec:poly} are based on kernels of size up to 
$(2N+1)\times(2N+1)$. This section aims at finding the 
value of the $N$ parameter that maximizes the security level.
For each approach, we have built 1,000 stego images with 
$N=2$, $4$, $6$, $8$, $10$, $12$, and $14$ where the covers belong 
to the BOSS contest database. 
This set contains 10,000 grayscale $512\times 512$ images in a RAW format. 
The security of the approach has been evaluated thanks to the Ensemble Classifier~\cite{DBLP:journals/tifs/KodovskyFH12} based steganalyser,
which is considered as a state of the art steganalyser tool.
This steganalysis process embeds the rich model (SRM) features~\cite{DBLP:journals/tifs/FridrichK12}
of size 34,671. 
For a payload $\alpha$, either equal to $0.1$ or to $0.4$, average testing errors (expressed in percentages) have been studied and are summarized in
Table~\ref{table:choice:parameter}.

\begin{table}[ht]
\caption{Average Testing Errors with respect to the the Kernel Size}
\begin{scriptsize}
\centering
\setlength{\tabcolsep}{3pt} 
\begin{tabular}{|c|c|c|c|c|c|c|c|c|}
\cline{2-9}
\multicolumn{1}{c|}{} & \multirow{2}{*}{$\alpha$} & \multicolumn{7}{c|}{$N$} \\
\cline{3-9}
\multicolumn{1}{c|}{}&  & $2$ & $4$&  $6$&  $8$& $10$& $12$ & $14$ \\
\hline{}
Average testing  
& \textit{0.1} & 39& 40.2&  39.7&  39.8& 40.1& $39.9$&  $39.8$  \\
\cline{2-9}
error for Kernel $K_y$ & \textit{0.4}& 15& 18.8& 19.1&  19.0& 18.6& 18.7 & 18.7 \\
\hline
Average testing & \textit{0.1} & 35.2 & 36.6&  36.7&  36.6& 37.1& 37.2 & 37.2 \\
\cline{2-9}
error for Kernel $K_o$ & \textit{0.4}  & 5.2 & 6.8& 7.5 & 7.9 & 8.1 & 8.2 & 7.6 \\
\hline
\end{tabular}
\label{table:choice:parameter}
\end{scriptsize}
\end{table}

Thanks to these experiments, we observe that the size $N=4$ (respectively $N=12$) obtains sufficiently  large average testing errors for the $Ky$ based approach 
(resp. for the $Ko$ based one). In what follows, these values are retained for these two methods.

\subsection{Security Evaluation}
As in the previous section, the BOSS contest database has been retained.
To achieve a complete comparison with other steganographic tools,
the whole database of 10,000 images has been used.
Ensemble Classifier with SRM features is again used to evaluate the security of the approach.

We have chosen 4 different payloads, 0.1, 0.2, 0.3, and 0.4, as in many steganographic evaluations.
Three values are systematically given for each experiment:
the area under the ROC curve (AUC),
the average testing error (ATE),
and the OOB error (OOB).

All the results are summarized in Table~\ref{table:experiments:summary}.
Let us analyse these experimental results. 
The security approach is often lower than those observed with state of the art tools:
for instance with payload $\alpha=0.1$, the most secure approach is WOW 
with an average testing error equal to 0.43 whereas our approach reaches 0.38.
However these results are  promising and for two reasons.
First, our approaches give more resistance towards Ensemble Classifier (contrary to HUGO)
for large payloads.
Secondly, without any optimisation, our approach is not so far from state of the art steganographic tools.
Finally, we explain the lack of security of the $Ko$ based approach with large payloads as follows:
second order derivatives are indeed directly extracted from polynomial interpolation.
This easy construction however induces large variations between the polynomial $L$ and 
the pixel function $P$.

\begin{table}
\caption{Summary of experiments}\label{table:experiments:summary}
\begin{small}
\centering
\begin{tabular}{|l|l|l|l|l|}
\hline

 & Payload & AUC & ATE   &  OOB \\ \hline
{WOW}   
            & 0.1 & 0.6501 & 0.4304 & 0.3974\\
            & 0.2 & 0.7583 & 0.3613 & 0.3169\\
            & 0.3 & 0.8355 & 0.2982 & 0.2488\\
            & 0.4 & 0.8876 & 0.2449 & 0.1978\\ 
                                       \hline

{SUNIWARD}  & 0.1 & 0.6542 & 0.4212 & 0.3972\\
            & 0.2 & 0.7607 & 0.3493 & 0.3170\\
            & 0.3 & 0.8390 & 0.2863 & 0.2511\\
            & 0.4 & 0.8916 & 0.2319 & 0.1977\\ 
 \hline
 {MVG}      & 0.1 & 0.6340 & 0.4310 &0.4124 \\
            & 0.2 & 0.7271 & 0.3726 &0.3399 \\
            & 0.3 & 0.7962 & 0.3185& 0.2858\\
            & 0.4 & 0.8486& 0.2719& 0.2353 \\ 
 \hline
  {HUGO}    & 0.1 & 0.6967 & 0.3982 & 0.3626 \\
            & 0.2 & 0.8012 & 0.3197 & 0.2847 \\
            & 0.3 & 0.8720 & 0.2557 & 0.2212 \\
            & 0.4 & 0.9517 & 0.1472 & 0.1230 \\ 
 \hline
{$Ky$ based approach} 
            & 0.1 & 0.7378 & 0.3768 & 0.3306 \\
            & 0.2 & 0.8568 & 0.2839 & 0.2408 \\
            & 0.3 & 0.9176 & 0.2156 & 0.1710 \\
            & 0.4 & 0.9473 & 0.1638 & 0.1324\\
 \hline
{$Ko$ based approach} 
            & 0.1 & 0.6831 & 0.3696  & 0.3450 \\
            & 0.2 & 0.8524 & 0.1302  & 0.2408 \\
            & 0.3 & 0.9132 & 0.1023  & 0.1045 \\
            & 0.4 & 0.9890 & 0.0880  & 0.0570 \\
                              \hline 
                              
\end{tabular}
\end{small}
\end{table}




\section{Conclusion}

The first contribution of this paper is to propose of a distortion
function which is based on second order derivatives. These
partial derivatives allow to accurately compute 
the level curves and thus to look favorably on pixels
without clean level curves. 
Two approaches to build these derivatives have been proposed.
The first one is based on revisiting kernels usually embedded 
in edge detection algorithms. 
The second one is based on the polynomial approximation
of the bitmap image.
These two methods have been completely implemented.
The first experiments have shown that the security level 
is slightly inferior the one of the most stringent approaches. These first promising results encourage us to deeply investigate this research direction. 

Future works aiming at improving the security of this proposal are planned as follows. The authors want first to focus on other approaches to provide second order derivatives with larger discrimination power.
Then, the objective will be to deeply investigate whether the H\"older norm is optimal when the objective is to avoid null second order derivatives, and to give priority to the largest second order values.$\rightarrow$

\section*{\uppercase{Acknowledgements}}
This work is partially funded by the Labex ACTION program (contract ANR-11-LABX-01-01). 
Computations presented in this article were realised on the supercomputing
facilities provided by the M\'esocentre de calcul de Franche-Comt\'e.

\bibliographystyle{apalike}
\bibliography{example}

\end{document}